# Environmental Effect on the UV Optical Absorption of Single-Walled Carbon Nanotubes

Yoichi Murakami[1*] and Shigeo Maruyama[2*]

[1] *Department of Chemical System Engineering, The University of Tokyo, Bunkyo-ku, Tokyo 113-8656, Japan*

[2] *Department of Mechanical Engineering, The University of Tokyo, Bunkyo-ku, Tokyo 113-8656, Japan*

Corresponding authors:

ymurak@chemsys.t.u-tokyo.ac.jp, maruyama@photon.t.u-tokyo.ac.jp

**Abstract**

We studied optical absorption of single-walled carbon nanotubes by varying the dielectric environment. For the two different components of the broad UV absorption feature conventionally referred to as the π-plasmon, we find that the component at 5.0 - 5.3 eV exhibits remarkable spectral changes, based on which we attribute this to a dipolar radial surface plasmon. However, the component at ~4.5 eV remains unchanged, raising a fundamental question as to its conventional attribution. We discuss its relation with the absorption feature at ~4.5 eV in graphite arising from an interband transition.

Single-walled carbon nanotubes (SWNTs) are a novel one-dimensional (1-D) material made of an $sp^2$-bonded wall one atom thick. Strong confinement (~ 1 nm) of charge carriers in SWNTs results in their unique optical properties being dominated by strongly-bound excitons (Coulomb-bound electron-hole pairs) [1-3], recognized by the sharp optical absorption and photoluminescence peaks in near-infrared (NIR) region [4]. Due to this aspect, SWNTs have been proposed for various optical applications such as NIR detectors and light emitters [5,6].

It is known that these sharp optical features accompany an optical absorption baseline extending from the UV region, where there is a broad and intense optical absorption peak in the 4 - 6 eV range that is typically attributed to a π-plasmon excitation [7-9]. It has been pointed out that the intensity ratio between the excitonic absorption and NIR baseline absorption increases as the degree of SWNT isolation is enhanced [10], although this mechanism has not yet been clarified. So far, several theoretical studies have investigated UV absorption in SWNTs [11-14], but many aspects are inconsistent and still under debate. Therefore, clear understanding of the UV absorption features is essential to fully utilize SWNTs for suggested optical applications.

Previously, we showed that the broad UV absorption peak in the 4 - 6 eV range actually consists of two distinct components [15,16]. In particular, the absorption feature at ~4.5 eV is observed for light polarized parallel to the SWNT axis, while the other absorption feature at 5.0 - 5.3 eV is observed for light polarized perpendicular to the axis [15]. These polarization dependences have recently been confirmed by optical absorption measurements of highly individualized SWNTs [17,18] and angle-resolved electron energy-loss spectroscopy (AR-EELS) of aligned SWNTs [19]. Also in our previous report, we attributed the parallel component (~4.5 eV) to the maximum of the imaginary part of the dielectric function perpendicular to the graphene *c*-axis (i.e., in-plane direction) Im{$\varepsilon_\perp$}, and the perpendicular component (5.0 - 5.3 eV) to the maximum of Im{$-\varepsilon_\parallel^{-1}$} parallel to the *c*-axis (out-of-plane direction)[15].

In this Letter, we show that the UV absorption feature in the 5.0 -5.3 eV range is sensitive to changes in the surrounding dielectric environment, while the other absorption feature at ~4.5 eV is unaffected. Based on these experimental results, we discuss the properties as well as physical origins of the two UV absorption features.

We used vertically-aligned SWNTs (VA-SWNTs) directly grown on a quartz substrate [20,21] by the alcohol chemical vapor deposition method [22] for the optical absorption measurements. Figure 1(a) shows a field-emission scanning electron microscope (FE-SEM) image of the sample. The thickness of the VA-SWNT film is typically 5 to 10 μm. The order parameter is known to be ~0.75 [15,16], which corresponds to an average deviation of 24° from ideal alignment. The SWNTs form thin bundles with diameters ≤10 nm [23]. Figure 1(b) shows a high-resolution transmission electron microscope (HR-TEM) image of VA-SWNTs transferred directly onto a TEM grid by a mechanical (dry) process. The image shows that the SWNTs are free from amorphous carbons and graphitic impurities. The high quality and purity of our VA-SWNTs has been confirmed by the high "G-to-D ratio" from Raman scattering measurements [20] and by the result of an "optical TGA" measurement where VA-SWNTs burned only at around 600 °C (Fig. 8 of Ref. 24). In the following, optical absorption spectra of VA-SWNTs were measured with light normally incident on the substrates, i.e., polarized perpendicular to the direction of ideal alignment of the VA-SWNTs.

Figure 2(a) compares two absorption spectra, the spectrum of the sample measured in air (black solid curve) and that of the same sample subsequently measured in acetonitrile (red dot-dash curve). Acetonitrile ($CH_3CN$) has a high dielectric constant ($\varepsilon \sim 38$) and high optical transparency up to 6 eV. The optical absorption in NIR was reduced by immersion of the sample into acetonitrile, and is explained by the change in the Fermi level caused by charge-transfer between SWNTs and the surrounding medium [25,26]. At higher energies, the magnitude of this difference becomes smaller until both spectra coincide for photon energies above 1.5 eV. However, the spectral change is again observed in the UV region around 5 eV, which was not known previously.

Figure 2(b) compares spectra measured from a VA-SWNT sample in a "desorbed" state (minimal molecular adsorption, black solid curve) and an "adsorbed" state (significant molecular adsorption, red dot-dash curve). The adsorbed state was achieved by storing the sample in a container filled with air for four months after its synthesis. The desorbed state was achieved by heating the same sample for 1 h at 200 °C in a low-pressure Ar atmosphere to desorb molecules from the SWNTs. The measurement of the adsorbed state was performed first, then heat treatment, which was immediately followed by measurement of the same "desorbed" sample. Figure 2(b) also shows reduction of the absorption intensity in NIR. Exposure of SWNTs to air is known to result in *p*-type doping, caused by adsorption of molecular oxygen and/or O-H species [27-28]. Similar to the case of acetonitrile immersion [Fig. 2(a)], a spectral change is also seen in UV. It is noted that these results have revealed that *only* the optical absorption feature at 5.0 - 5.3 eV shows a spectral change, while that at ~4.5 eV does not.

We show that the same optical change can be induced by a simple electro-chemical method. The measurement setup consists of a VA-SWNT substrate (working electrode, W. E.), a Pt wire (counter electrode, C. E.), and an Ag wire (reference electrode, R. E.), set in an optically-transparent quartz cell filled with acetonitrile and 0.1 M lithium perchlorate ($LiClO_4$), which served as an electrolyte. The voltage was applied between the W. E. and C. E., while the magnitude of the applied voltage was measured between the W. E. and R. E. Figure 3(a) shows the change in optical absorption as the voltage was varied from 0.25 to 1.0 V. Figure 3(b) shows the subsequent reverse process (1.0 to 0.25 V). In addition to the expected reduction of optical absorption below 2 eV, we also see a change in the 5.0 - 5.3 eV region. Figure 3(c) plots the induced changes at 1.3 and 5.1 eV. Each measurement was taken approximately 5 min after each voltage change. For 5.1 eV, the last measurement at 0.25 V (marked with an asterisk) was taken 15 min after the previous measurement at the same voltage, indicating a much slower response for the optical change at 5.1 eV than that at 1.3 eV.

As shown in Figs. 2 and 3, the perpendicular component (5.0 - 5.3 eV) is influenced by changes in the environment. In the case of 1-D metallic nanowires, it is known that the direction perpendicular to the nanowire axis (the direction of electron confinement) is the only direction in which dipolar surface plasmons can be excited [29,30]. Figure 4 shows schematics of dipolar surface plasmons induced on the surface of cylindrical nanowire (viewed along the axis direction) with complex dielectric function $\varepsilon$. The applied external field and dielectric function of the surrounding medium are denoted by $\boldsymbol{E}_0$ and $\varepsilon_m$, respectively. The depolarization field $\boldsymbol{E}_{dep}$, created by induced charges on the interface is given by [29]

$$\boldsymbol{E}_{\text{dep}} = -\frac{\varepsilon - \varepsilon_{\text{m}}}{\varepsilon + \varepsilon_{\text{m}}} \boldsymbol{E}_0 . \qquad (1)$$

The resultant field inside the nanowire $\boldsymbol{E}_{\text{in}}$, proportional to optical absorption intensity, is

$$\boldsymbol{E}_{\text{in}} = \boldsymbol{E}_0 + \boldsymbol{E}_{\text{dep}} = \frac{2\varepsilon_{\text{m}}}{\varepsilon + \varepsilon_{\text{m}}} \boldsymbol{E}_0 , \qquad (2)$$

which is sensitive to the value of $\varepsilon_{\text{m}}$. Furthermore, the photon energy corresponding to maximum dipolar surface plasmon excitation depends on both $\varepsilon_{\text{m}}$ and the charge density (or plasma frequency $\omega_{\text{p}}$) of the material [29].

Changes observed in the perpendicular component are well explained in terms of dipolar surface plasmons. The significant enhancement of this absorption [Fig. 2(a)] is explained by the much larger $\varepsilon_{\text{m}}$ of acetonitrile than of air [Eq. (2)]. The weakening and the red shift of this feature observed in Fig. 2(b) may be ascribed to *p*-type doping (withdrawal of electrons from SWNTs) caused by the adsorption of oxygen molecules [27,28], as well as a change in $\varepsilon_{\text{m}}$ due to the adsorbed molecules. More importantly, the difference in the observed transient changes of the absorbance at 1.3 and 5.1 eV [Fig. 3(c)] suggest that the dominant mechanisms responsible for the optical changes at these energies are different. The former is caused by the change in the Fermi level, which always results in the reduction of absorption intensities regardless of *p*- or *n*-type doping [25]. On the other hand, the much slower change observed at 5.1 eV suggests that this feature is influenced by a slower process, such as diffusion of the electrolyte around the SWNTs. This is plausible considering that 5.1 eV is considerably far from the Fermi level and no optical changes were observed in the intermediate (2 - 4 eV) region.

On the other hand, no change was observed at ~4.5 eV (parallel component) by the change of environment. It is noted that this feature is, at least, not a dipolar surface plasmon because this direction is not the direction of confinement. So far, the parallel component at ~ 4.5 eV has been regarded as a π-plasmon [7-9]. However, in the case of SWNTs *all atoms constitute the surface*, thus the π-electrons are directly exposed to the surrounding medium. If this feature truly originated from a π-plasmon, it is expected to be affected by the change in dielectric environment. Since this is not observed, our experimental results raise a fundamental question as to the physical origin of this parallel component of UV absorption. A recent AR-EELS study of the same VA-SWNTs has also recognized the peak component at ~4.5 eV by extrapolating the obtained EELS spectra to the case of zero momentum transfer ($q \rightarrow 0$, analogous to optical excitation) [19]. However, we note that the observation of a peak in EELS is not sufficient to conclude that the peak arises from plasmons, because *interband electron transitions can also be excited in EELS* (although as an indirect transition accompanying nonzero $q$) [31].

In the case of graphite, the origin of the strong UV absorption peak at ~4.5 eV has been recognized as a *π → π\* interband electronic transition* at the M (also called Q) point of the Brillouin Zone, corresponding to the maximum of Im$\{\varepsilon_\perp\}$ [32-35]. One of the most essential differences between SWNTs and graphite/graphene is their *dimension*. In a pure 1-D system, it is known that only collective excitations can exist [36]. Based on this view point, electronic excitations in SWNTs are essentially collective as long as SWNTs are regarded as a 1-D system. In this context, we must address the issue of terminology, i.e., what to call the UV absorption feature at ~4.5 eV in SWNTs. However, apart from semantics, it is more important to clarify its physical origin. Further theoretical investigations are required to conclude the issues discussed above, especially regarding the origin of the parallel UV absorption feature of SWNTs observed at ~4.5 eV.

In summary, we studied the two UV absorption features of SWNTs in the 4 - 6 eV range that have so far both been referred to as π-plasmons. We found that changes in the dielectric environment induce remarkable changes in the perpendicular component at 5.0 - 5.3 eV. This feature is explained as a radial dipolar surface plasmon induced in SWNTs. On the other hand, we found that the parallel UV absorption component at ~4.5 eV was unaffected by the environmental changes. This raises a fundamental question as to the physical origin of this absorption feature, which has heretofore been classified as a π-plasmon. We anticipate that the findings and discussions presented in this paper will stimulate further investigations and elucidation of the properties of UV absorption in SWNTs.

**Acknowledgement**

We thank Erik Einarsson for refinement of the manuscript. Part of this work was financially supported by Grant-in-Aid for Scientific Research (19206024 and 19054003) from the Japan Society for the Promotion of Science, NEDO (Japan), and MITI's Innovation Research Project on Nanoelectronics Materials and Structures. Y. M. was financially supported by the JSPS grant #18-09883.

**Figure Captions**

FIG. 1: (a) FE-SEM image of VA-SWNTs grown on the surface of a quartz substrate. (b) HR-TEM image of VA-SWNTs directly transferred onto a TEM grid.

FIG. 2: (Color online) (a) Comparison of the optical absorption spectra of the same VA-SWNT sample measured in air (black solid) and subsequently in acetonitrile (red dot-dash). Spikes below 0.8 eV are absorptions by acetonitrile. (b) Comparison of the optical absorption spectra of an identical VA-SWNT sample measured in a molecular-adsorbed state (red dot-dash) and after desorption of the molecules (black solid). In panels (a) and (b), green dashed curves show absolute values of the differences between the two spectra, and vertical dotted lines indicate the positions of the two distinct UV absorption features. All spectra are shown *as measured*, i.e., no scaling operations have been performed.

FIG. 3: (Color online) (a, b) Electrochemically induced changes in the absorption spectra measured in 0.1 M LiClO$_4$ in acetonitrile. The voltage was changed from 0.25 to 1.0 V [panel (a)], and subsequently from 1.0 to 0.25 V [panel (b)]. (c) Absorbance changes measured at 1.3 eV (black squares) and 5.1 eV (red circles). Open symbols connected by solid lines are for the 0.25 → 1.0 V process, while filled symbols connected by dotted lines are for 1.0 → 0.25 V. The last point for 5.1 eV at 0.25 V (marked by an asterisk) was measured 15 min after the previous measurement at the same voltage. Three spectra shown in panels (a) and (b) are from the experiment in panel (c), corresponding to 0.25, 0.6, and 1.0 V.

FIG. 4: (Color online) Schematics of dipolar surface plasmons induced on the surface of cylindrical nanowires by an external electric field $\boldsymbol{E}_0$, viewed along the axial direction. $\varepsilon$, $\varepsilon_m$: Complex dielectric functions of the nanowire and surrounding media, respectively. $\boldsymbol{E}_{dep}$, $\boldsymbol{E}_{in}$: Depolarization electric field and the resultant field inside the nanowire, respectively. Panels (a) and (b) depict $\varepsilon_m > \varepsilon > 0$ and $\varepsilon > \varepsilon_m > 0$ cases, respectively.

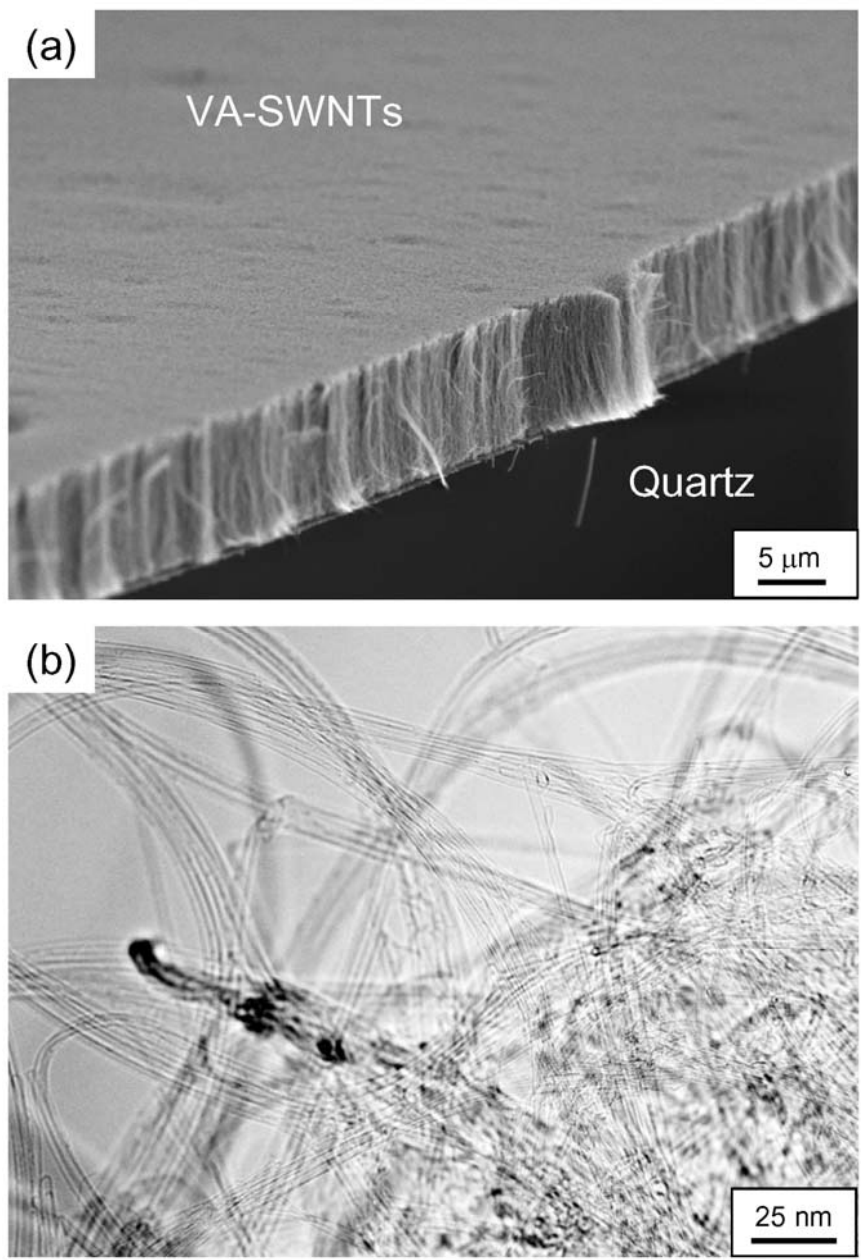

Figure 1

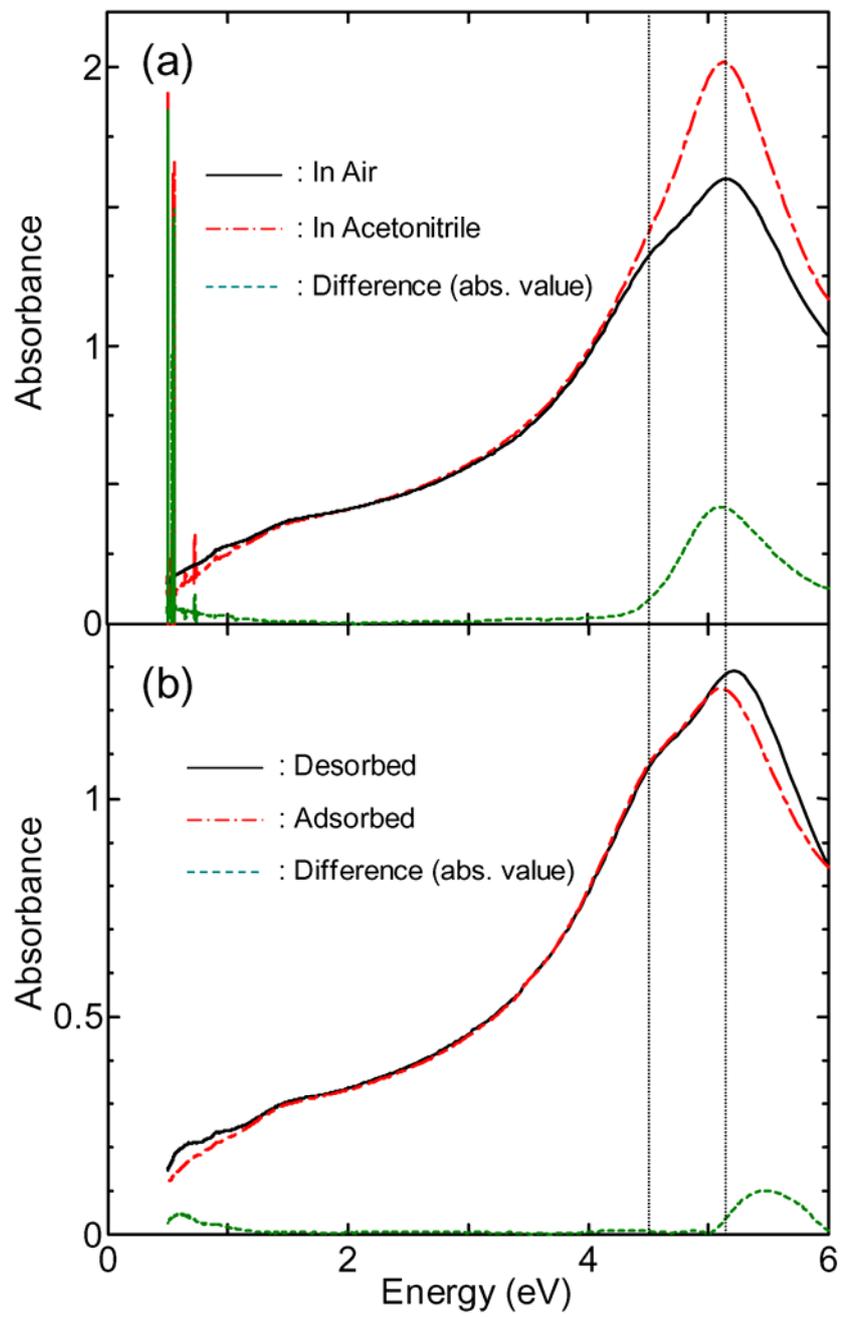

Figure 2

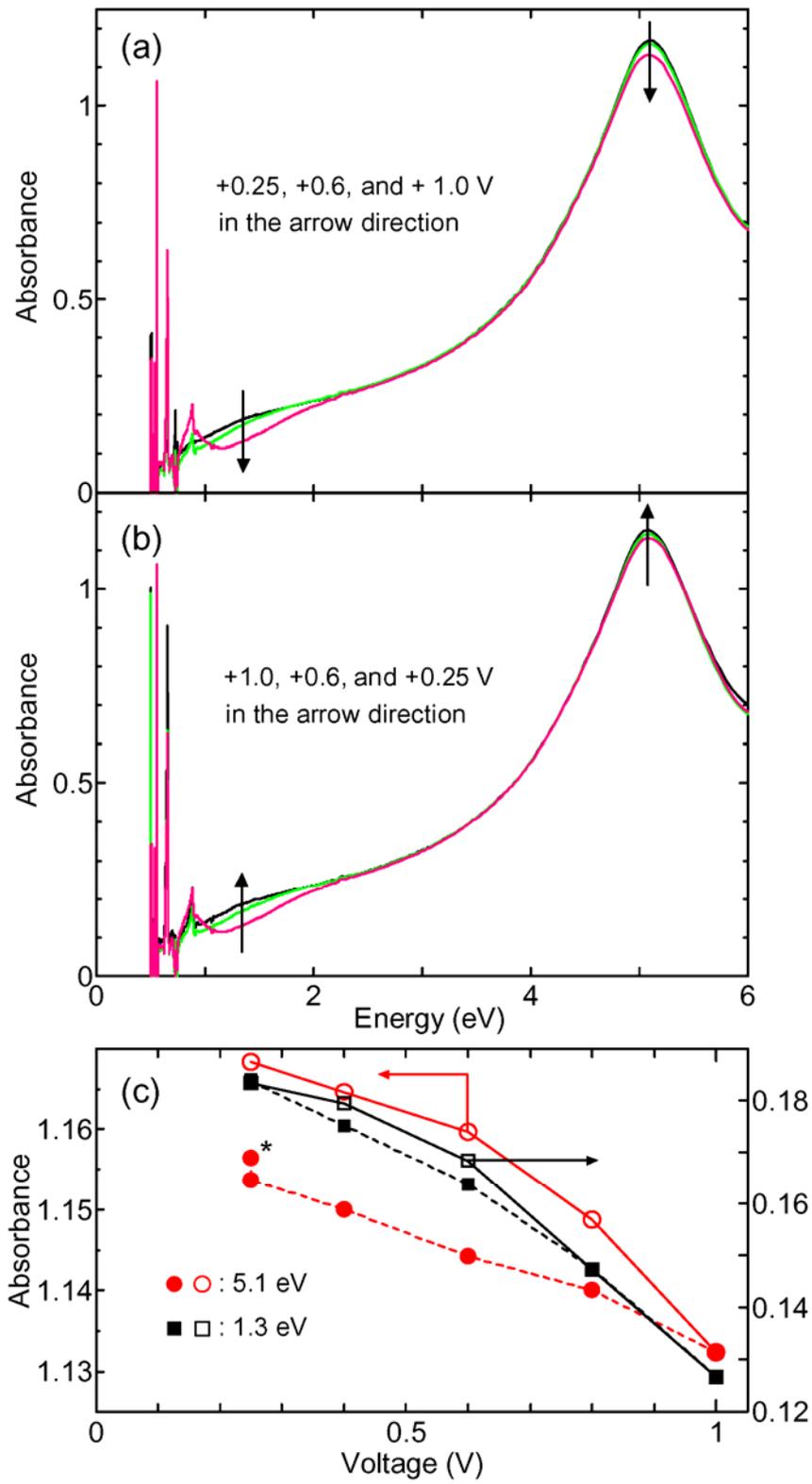

Figure 3

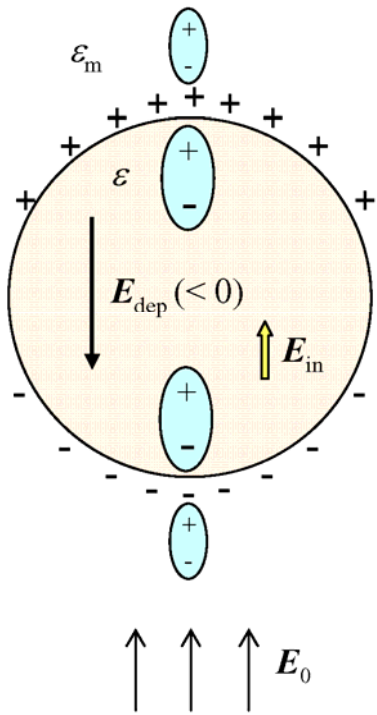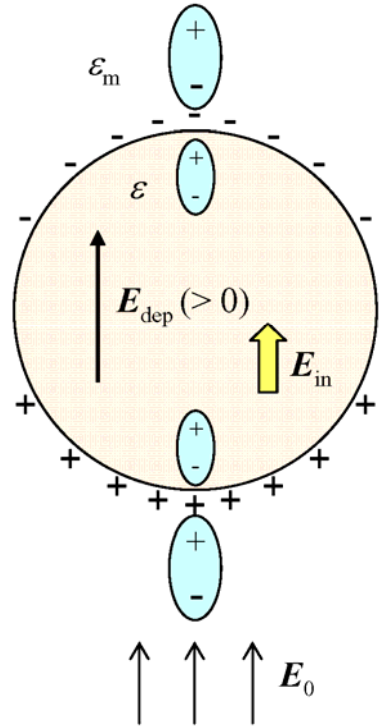

Figure 4